\documentclass[preprint2]{aastex701}
\begin{document}

\title{The potential broader complex linked to the key Cepheid SV Vul}

\author[orcid=0000-0001-8803-3840]{Daniel Majaess}
\affiliation{Mount Saint Vincent University, Halifax, Nova Scotia, Canada.}
\email[show]{Daniel.Majaess@msvu.ca}

\author[0000-0003-1952-3680]{Ignacio Negueruela}
\affiliation{Departamento de Fisica Aplicada, Facultad de Ciencias, Universidad de Alicante, Carretera de San Vicente s/n, E-03690 San Vicente del Raspeig, Spain}
\email[]{ignacio.negueruela@ua.es}

\author[]{Leonid N. Berdnikov}
\affiliation{Sternberg Astronomical Institute, Moscow State University, Universitetskii Pr. 13, Moscow 119992, Russia}
\email[]{lberdnikov@yandex.ru}

\author[0000-0002-4102-1751]{Charles J. Bonatto}
\affiliation{Departamento de Astronomia, Universidade Federal do Rio Grande do Sul, CP 15051, RS, Porto Alegre 91501-970, Brazil}
\email[]{charles.bonatto@ufrgs.br}

\author[0000-0003-1184-1860]{David G. Turner}
\affiliation{Saint Mary's University, Halifax, Nova Scotia, Canada}
\email[]{David.Turner1@smu.ca}

\author[0000-0002-7064-099X]{Dante Minniti}
\affiliation{Instituto de Astrofísica, Dep.~de Física y Astronomía, Facultad de Ciencias Exactas, Universidad Andres Bello, Av.~Fernández Concha 700, Santiago, Chile}
\affiliation{Vatican Observatory, Specola Vaticana, V-00120, Vatican City, Vatican City State}
\email[]{vvvdante@gmail.com}

\author[0000-0003-1184-1860]{Vittorio F. Braga}
\affiliation{INAF - Osservatorio Astronomico di Roma, Via di Frascati 33, 00078 Monte Porzio Catone, Italy}
\email[]{vittorio.braga@inaf.it}

\author[0000-0002-0155-9434]{Giovanni Carraro}
\affiliation{Dipartimento di Fisica e Astronomia “Galileo Galilei,” Università degli Studi di Padova, Vicolo Osservatorio 3, I-35122, Padova, Italy}
\email[]{giovanni.carraro@unipd.it}

\author[0009-0009-8613-0087]{Igor Usenko}
\affiliation{Astronomical Observatory, Odessa National University, Marazliivska 1-B, Odessa, 65014, Odessa, Ukraine}
\affiliation{Mykolaiv Astronomical Observatory Research Institute, Obsevatorna 1, Mykolaiv, 54030, Mykolaiv, Ukraine}
\email[]{igus99@ukr.net}

\author[0000-0002-4896-8841]{Giuseppe Bono}
\affil{Department of Physics, Università di Roma Tor Vergata, via della Ricerca Scientifica 1, 00133, Rome, Italy}
\affiliation{INAF - Osservatorio Astronomico di Roma, Via di Frascati 33, 00078 Monte Porzio Catone, Italy}
\email[]{bono@roma2.infn.it}

\author[0000-0002-4430-9427]{Matias Gomez}
\affiliation{Instituto de Astrofísica, Dep.~de Física y Astronomía, Facultad de Ciencias Exactas, Universidad Andres Bello, Av.~Fernández Concha 700, Santiago, Chile}
\email[]{matiasgomezcamus@gmail.com}

\author[0000-0001-6878-8648]{Roberto K. Saito}
\affiliation{Departamento de Física, Universidade Federal de Santa Catarina, Trindade 88040-900, Florianópolis, Brazil}
\email[]{roberto.saito@ufsc.br}

\author[0000-0002-4896-8841]{Maria G. Navarro}
\affiliation{INAF - Osservatorio Astronomico di Roma, Via di Frascati 33, 00078 Monte Porzio Catone, Italy}
\email[]{maria.navarro@inaf.it}

\begin{abstract}
A new distance was established to the Cepheid SV Vul ($P\simeq45^{d}$) and its host cluster Alicante 13, and is tied in part to deeper UKIDSS-DR6 photometry and Gaia DR3 observations ($d=2.30\pm0.13$ kpc). SV Vul and Alicante 13 possibly belong to a broader coeval stellar complex, which could include Liu-Pang 1738.  The clusters and Cepheid share comparable astrometry (e.g., $\sigma_\pi\simeq0.02$ mas, $\sigma_{\mu_\delta}\simeq0.06$ mas yr$^{-1}$), and similar cluster turnoffs (ages) exist as indicated by ultraviolet UVEX color-color analyses, Gaia XP spectroscopically differentially dereddened color-magnitude diagrams, and Padova isochrones. Yet SV Vul may be comparatively overluminous. Robust radial velocities for both clusters could substantiate certain hypotheses.  
\end{abstract}


\section{Introduction}
\citet{tur84} aimed to identify early-type stars associated with the classical Cepheid SV Vul ($P\simeq45^{d}$), with one objective being to secure calibrators for Cepheid relations via cluster membership.  The differentially reddened region features numerous Galactic associations and clusters \citep[\S5 in][]{neg20}, and is viewed nearly lengthwise through a spiral arm.  Three decades later the astrometry became available to facilitate disentangling phenomena along this complex sightline.  Specifically, \citet{neg20} relied on Gaia DR2 and high-resolution spectroscopy to discover the Cepheid's host cluster (Alicante 13, $d\simeq 2.5$ kpc). 

\begin{figure*}[t]
\begin{center}
 \includegraphics[width=2.0\columnwidth]{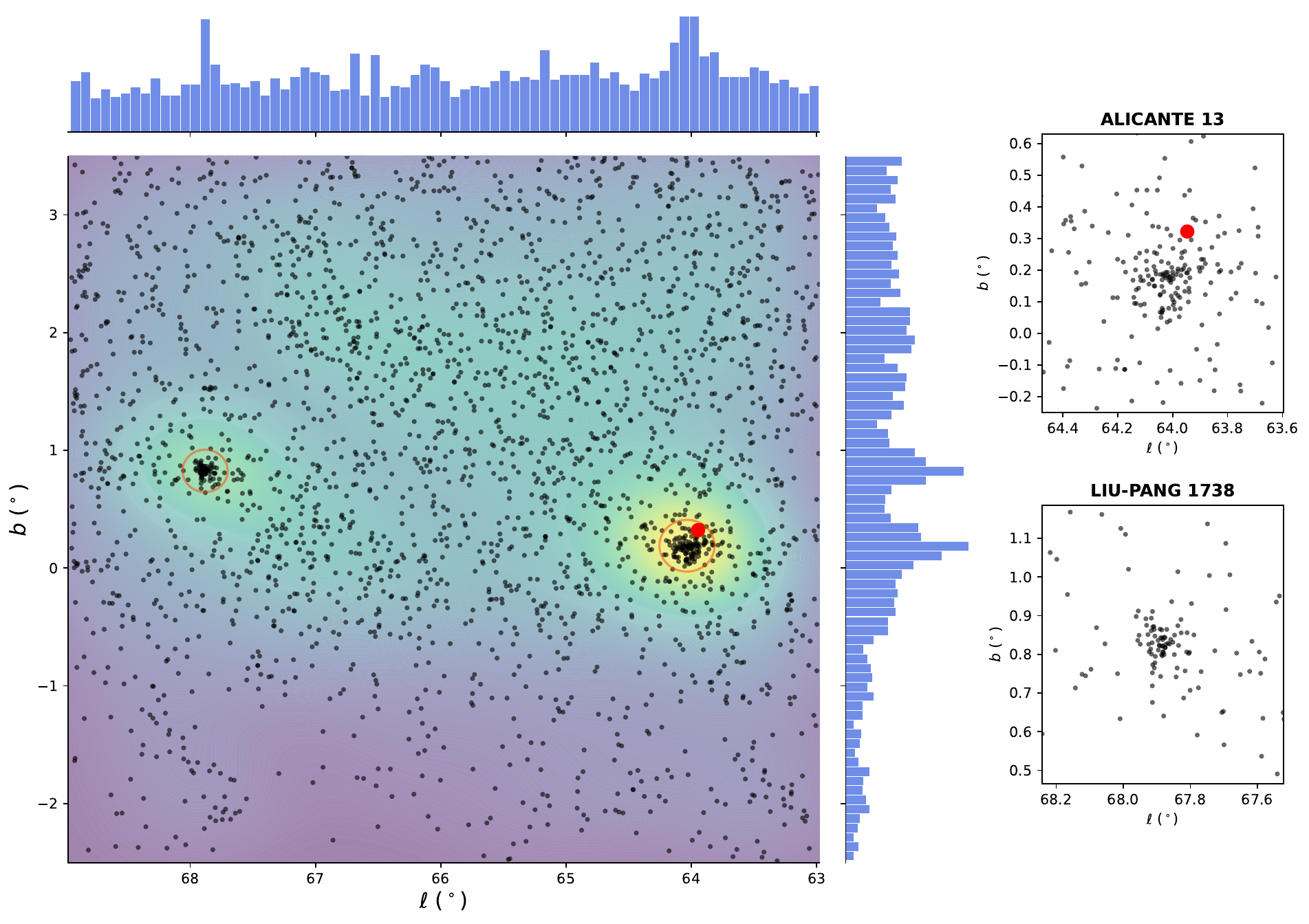}
  \caption{Stars in the region exhibiting similar Gaia DR3 astrometry (\S \ref{sec:astrometry}). The two principal overdensities are Liu-Pang 1738 (left) and Alicante 13 (right), and SV Vul is the red datum. The clusters possess similar turnoffs (ages) and astrometry (Table~\ref{table:parameters}), such as $\pi= 0.395\pm0.021$ and $0.427\pm0.023$ mas ($0.402\pm0.021$ mas for SV Vul). Parallaxes were corrected for the floating DR3 zeropoint following \citet{lin21}.  The encompassing circles are tied to $rmax$ from \citet{lp19}.}
 \label{fig-pm}
\end{center}
\end{figure*}

This study aims to build upon prior research on SV Vul and the encompassing region \citep{tur84,neg20}, namely by holistically relying on new and unused deeper observations \citep[e.g., DR3,][]{gaia22}. The analysis clarifies potential connections in the broader field, provides desirable redundancies relative to the derivation of cluster parameters, and synthesizes diverse evidence regarding SV Vul's intrinsic brightness.  Gaia DR3 astrometry ($\pi$, $\mu_{\alpha}$, $\mu_{\delta}$), photometry ($B$, $G$, $R$), XP spectra ($E(B-R)$, $A_G$), and faint ultraviolet UVEX observations ($u$, $g$, $r$) are mobilized to produce color-color and differentially dereddened color-magnitude diagrams.  Moreover, deeper UKIDSS-DR6 photometry is utilized to create near-infrared color-magnitude diagrams ($J$, $H$).

\section{Analysis}
\subsection{Astrometry}
\label{sec:astrometry}
The python \texttt{DBSCAN} package was employed to characterize cluster astrometry (e.g., $n=15$ member minimum).  Deduced formal parameters for Alicante 13 are $\pi=0.397\pm0.023$ mas and $\mu_{\alpha},\mu_{\delta}=(-2.102\pm0.096,-5.894\pm0.130$) mas yr$^{-1}$ ($n=152$). A preliminary search for a broader complex was undertaken by expanding parallax and proper motion uncertainties by $5$ and $2\sigma$ to maximize sampling, respectively \citep[e.g., to include Alicante 13 members BD+273542, HD 339063, HD 339064,][]{neg20}.  Those aforementioned evolved stars are $<1.5\arcmin$ from the cluster center, exhibit comparable DR3 astrometry within expected intrinsic cluster deviations, and possess independent spectroscopy and velocities \citep[][]{neg20}.  Fig.~\ref{fig-pm} unveils two principal overdensities, whereby the proximate cluster to SV Vul (red datum) is Alicante 13. Gaia DR3 astrometry for SV Vul is $\pi=0.373\pm0.021$ mas and $\mu_{\alpha},\mu_{\delta}=(-2.158\pm0.016,-5.962\pm0.021)$ mas yr$^{-1}$.  The other main cluster in Fig.~\ref{fig-pm} was characterized by \citet[][DR2, ID=1738]{lp19}.  

For Liu-Pang 1738 the \texttt{DBSCAN} findings are $\pi=0.366\pm0.022$ mas and $\mu_{\alpha},\mu_{\delta}=(-2.168\pm0.090,-6.014\pm0.115)$ mas yr$^{-1}$ ($n=80$). The \citet[][L21]{lin21} DR3 corrected parallaxes for Liu-Pang 1738 (UBC 134), Alicante 13 (UBC 130), and SV Vul are: $\pi=0.395\pm0.021,0.427\pm0.023,0.402\pm0.021$ mas, accordingly. Admittedly, there are varied opinions regarding the DR3 zeropoint \citep[e.g., Fig.~10 in][]{mol23}. By comparison, using DR2 \citet[][]{can20} determined $\pi=0.393\pm0.030$ mas (Alicante 13), and \citet[][]{lp19} obtained $\pi=0.367\pm0.022$ mas (Liu-Pang 1738). The results presented here and in the literature are comparable (Table~\ref{table:parameters}).

\subsection{Ultraviolet color-color diagram}
\begin{figure*}[t]
\begin{center}
 \includegraphics[width=1\columnwidth]{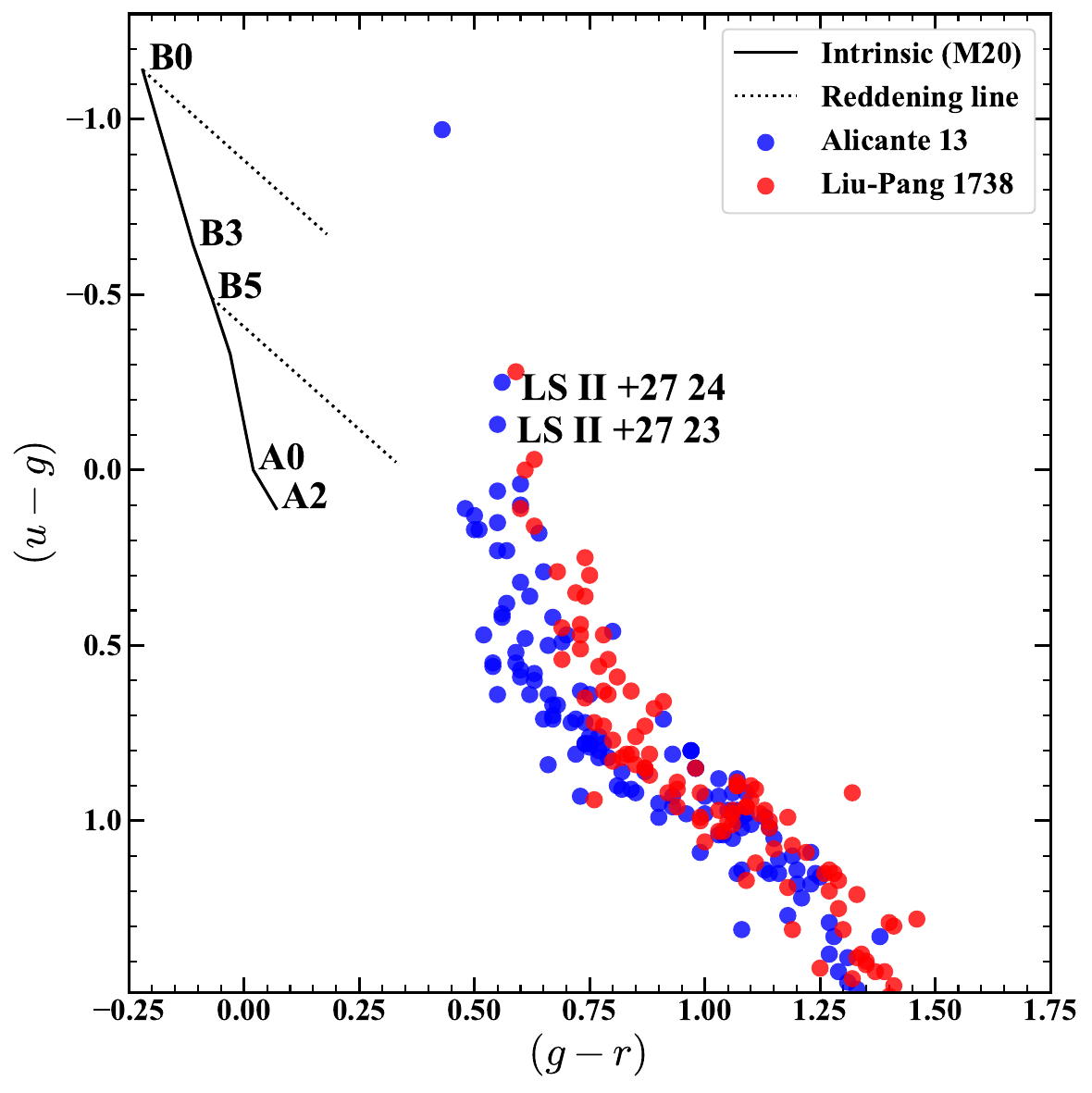}
 \includegraphics[width=1\columnwidth]{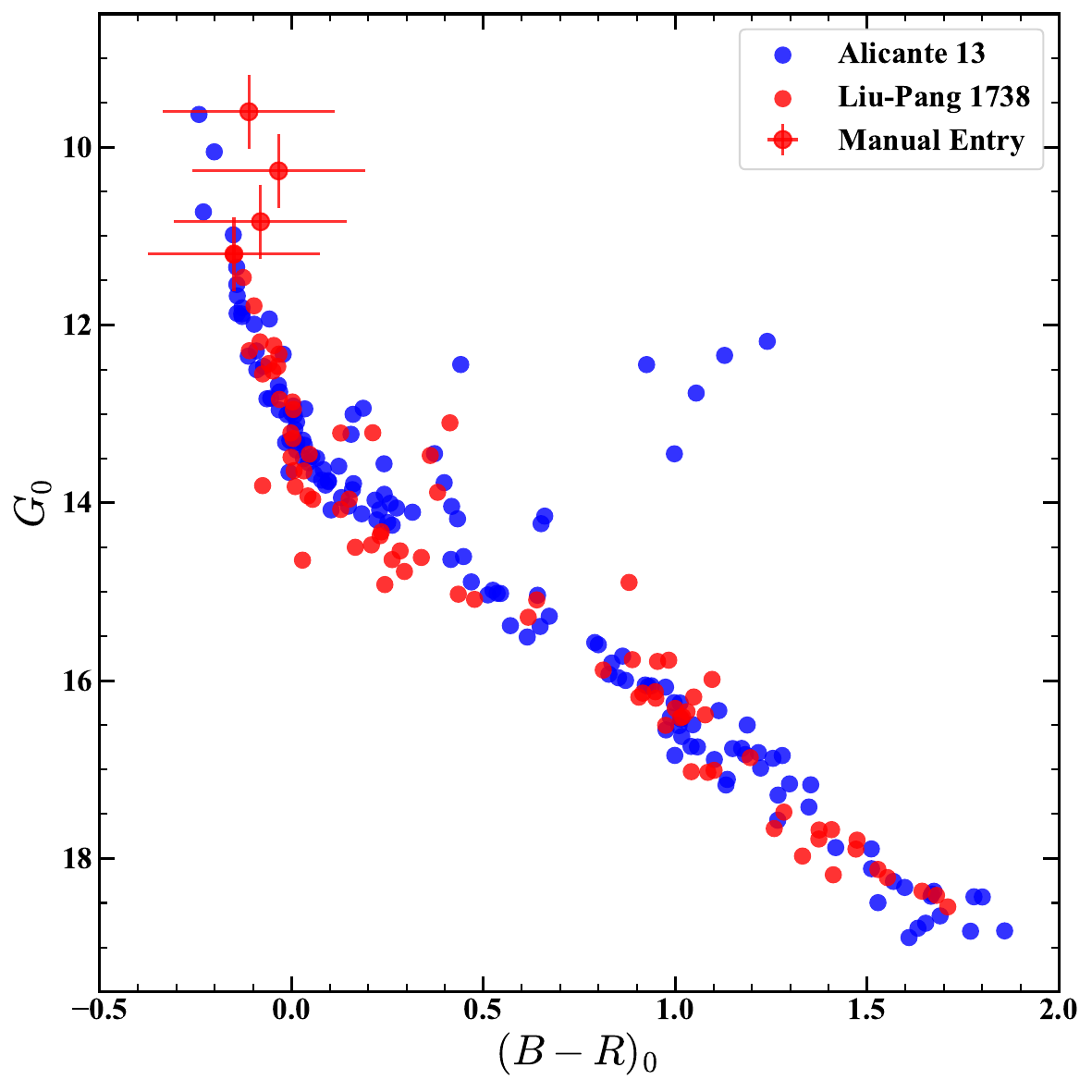}
  \caption{Left, an ultraviolet UVEX color-color diagram conveys that Alicante 13 and Liu-Pang 1738 possess similar turnoffs.  Two B2.5 stars from \citet[][]{neg20} are annotated. Liu-Pang 1738 is observed through additional obscuration. Right, a differentially dereddened color-magnitude diagram constructed using published Gaia XP spectroscopic parameters for a subset of DR3 stars.  The clusters appear coeval within uncertainties, with Liu-Pang 1738 being slightly more remote. Four bright Liu-Pang 1738 stars were manually added owing to spurious parameters or absence from the initial XP release (\S \ref{sec:dr}).}
 \label{fig-ccd}
\end{center}
\end{figure*}

\begin{figure*}[t]
\begin{center}
 \includegraphics[width=1\columnwidth]{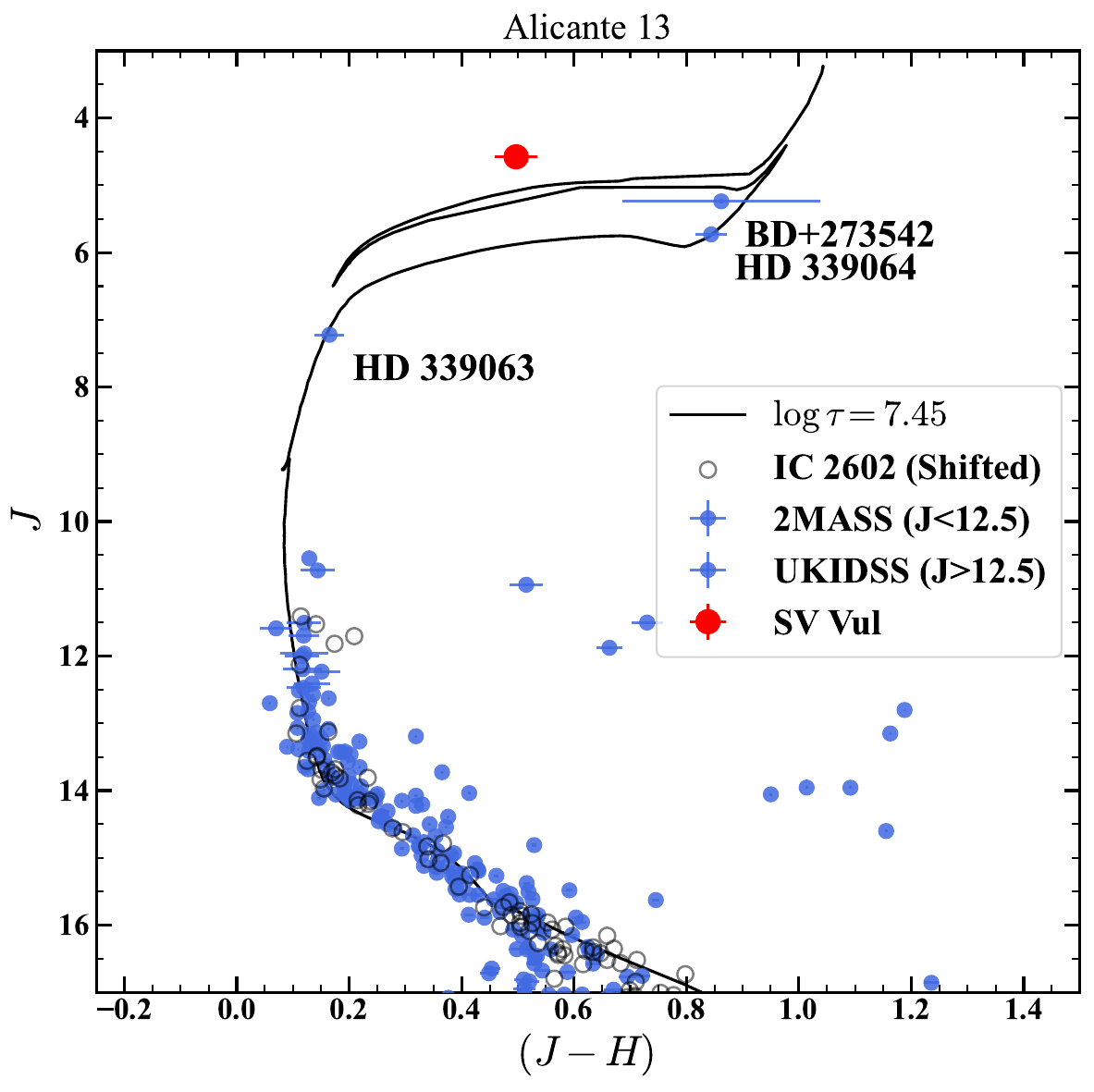}
 \includegraphics[width=1\columnwidth]{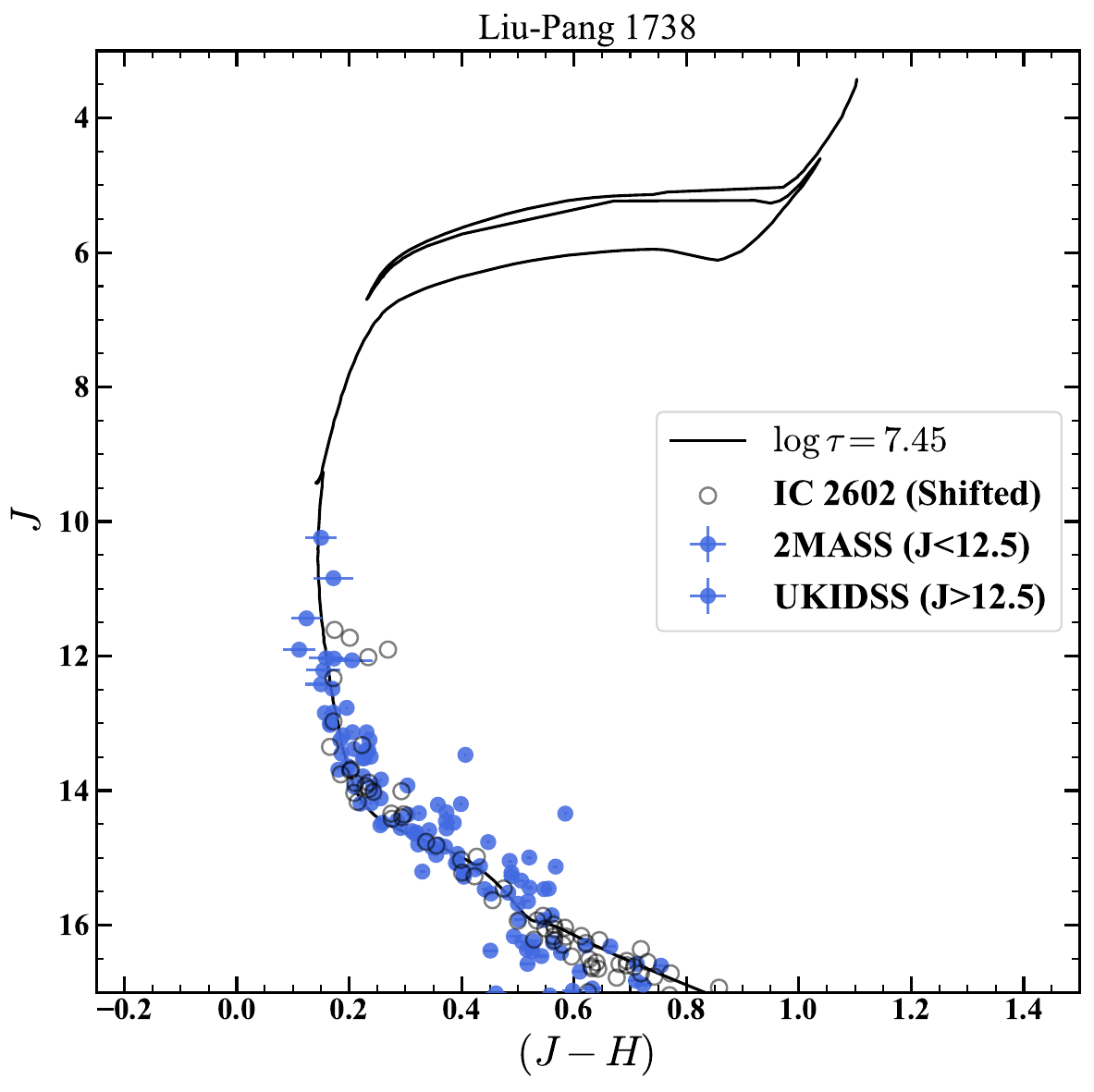}
  \caption{2MASS-UKIDSS color-magnitude diagrams for Alicante 13 and Liu-Pang 1738 ($d=2.30, 2.40(\pm0.13)$ kpc). Evolved stars BD+273542, HD 339063, and HD 339064 are highlighted (\S \ref{sec:astrometry}). Distances were derived by aligning a nearby unreddened and similarly young open cluster (IC 2602, open circles). A solar Padova PARSEC 1.2S-YBC $\log{\tau}=7.45\pm0.10$ isochrone was deduced from a visual fit, and is comparable to a mean inferred for SV Vul from Cepheid period-age relations ($\log{\tau}=7.35\pm0.17$). The isochrone was likewise overlayed upon Liu-Pang 1738 observations, since a similar age is supported by multiple indicators (e.g., Fig.~\ref{fig-ccd}). A semi-automated approach to inferring cluster parameters from numerous Padova models is conveyed in \S \ref{sec:redundancies}.  SV Vul appears overluminous relative to canonical isochrones (\S \ref{sec:svvul}).}
 \label{fig-cmd}
\end{center}
\end{figure*}

A scenario where the clusters are unrelated is disfavored by Alicante 13 and Liu-Pang 1738 displaying similar astrometry and turnoffs (i.e., comparable ages, Fig.~\ref{fig-ccd} and \S \ref{sec:redundancies}). The ultraviolet $ugr$ color-color diagram photometry stemmed from the UVEX-IGAPS surveys \citep[][]{mon20}, and the intrinsic main-sequence relation shown in Fig.~\ref{fig-ccd} is likewise from \citet{mon20}.  A linear $E(u-g)/E(g-r)=1.17\pm0.14$ vector was estimated using the python \texttt{EXTINCTION} routine \citep[e.g.,][]{odo94}, together with passband information described in \citet{mon20}. The selected total to selective extinction ratio is $R_V=3.1$, though that value is debated whereby for example \citet[][]{ber96} deduced $R_V\simeq 3.26$ using Cepheids. For Figs.~\ref{fig-ccd} and \ref{fig-cmd} the astrometric criteria were again expanded to maximize membership (\S \ref{sec:astrometry}). \citet{lp19} estimated $rmax\simeq0.18, 0.22 ^{\circ}$ (max radii) for Liu-Pang 1738 and Alicante 13, respectively, and those cluster bounds were adopted for those figures.   

\citet[][their \S8.6]{mon20} underscored that standardizing terrestrial ultraviolet photometry is a longstanding challenge, which implies the data should be interpreted cautiously. As a result inferences emerging from Fig.~\ref{fig-ccd} are restricted to the following: a comparable turnoff occurs near early-B for Alicante 13 and Liu-Pang 1738, as inferred from the apparent photometry.  Two B2.5 stars from \citet[][]{neg20} are highlighted on the diagram (independent spectroscopy), and the A0 clamping point is discernible.   A robust determination of the turnoff spectral type solely from the diagram is hindered by a floating $u$-band zeropoint in concert with uncertainties associated with the reddening vector and intrinsic relation, yet redundancies are in place (e.g., \S \ref{sec:dr}, \S \ref{sec:redundancies}, and through spectroscopy).

\begin{deluxetable*}{lcccc}
\tablecaption{Summary of established parameters.\label{table:parameters}}
\tablehead{\colhead{} & \colhead{Alicante 13} & \colhead{Liu-Pang 1738} & \colhead{SV Vul} & \colhead{Ref.}}
\startdata
$\pi$ (mas) & $0.427\pm0.023$ & $0.395\pm0.021$ & $0.402\pm0.021$ & 1 (L21-DR3) \\
$\pi$ (mas) & $0.393\pm0.030$ & $0.368\pm0.029$ & $0.373\pm0.030$ & 2 (DR2) \\
$\mu_{\alpha}$ (mas yr$^{-1}$) & $-2.102\pm0.096$ & $-2.168\pm0.090$ & $-2.158\pm0.016$ & 1 (DR3)\\
$\mu_{\alpha}$ (mas yr$^{-1}$) & $-2.099\pm0.073$ & $-2.174\pm0.052$ & $-2.139\pm0.045$ & 2 (DR2) \\
$\mu_{\delta}$ (mas yr$^{-1}$) & $-5.894\pm0.130$ & $-6.014\pm0.115$ & $-5.962\pm0.021$ & 1 (DR3) \\
$\mu_{\delta}$ (mas yr$^{-1}$) & $-5.856\pm0.064$ & $-5.983\pm0.098$ & $-5.820\pm0.050$ & 2 (DR2) \\
$d$ (kpc) & $2.342^{+0.133}_{-0.120}$ & $2.532^{+0.142}_{-0.128}$ & $2.488^{+0.137}_{-0.124}$ & 1 (L21-DR3) \\
$d$ (kpc) & $2.30\pm0.13$ & $2.40\pm0.13$ & $-$ & 1 \\
& $2.30\pm0.12$ & $2.49\pm0.11$ & $-$ & 1 \\
$\log{\tau}$ & $7.53\pm0.05$ & $7.60\pm0.04$ & $7.35\pm0.17$ & 1 \\
& $7.44, 7.48$ & $7.341\pm0.287$, $7.87$ & $-$ & 2,3,4,2 \\
\enddata
\tablenotetext{ }{Notes:~references pertain to research presented here (1), \citealt{can20} (2), \citealt{neg20} (3), and \citealt{dia21} (4).}
\end{deluxetable*}

\subsection{Differentially dereddened optical color-magnitude diagram}
\label{sec:dr}
The common turnoff inferred from the UVEX ($u-g$)/($g-r$) color-color analysis was bolstered by a differentially dereddened Gaia $G_0-(B-R)_0$ diagram (Fig.~\ref{fig-ccd}). Extinction parameters $A_G$ and $E(B-R)$ were published for a subset of Gaia DR3, and were inferred from model fits to low-resolution XP spectra ($\lambda \simeq 330-1050$ nm).  For example, certain data are associated with MARCS and PHOENIX models.  Concerns persist regarding systematics within this first XP release \citep[e.g.,][and discussion therein]{and23}, and \citet{mt24} identified discernible shifts between DR3 spectroscopically dereddened main-sequences and unobscured clusters.  Nonetheless, the dereddened color-magnitude plot confirms Liu-Pang 1738 and Alicante 13 are of comparable age to within the uncertainties, with Alicante 13 possibly being younger (Fig.~\ref{fig-ccd}, see also Table~\ref{table:parameters}). Moreover, Liu-Pang 1738 appears more distant.  

Systematics endemic to the XP release would improbably overlay Alicante 13 upon Liu-Pang 1738, and nevertheless redundancies exist via the color-color diagram and isochrones (Table~\ref{table:parameters}).  XP spectra can discern clusters markedly offset in age, as conveyed between NGC 7788 and the potential binary Cepheid cluster NGC 7790/Berkeley 58 \citep[left panel of Fig.~2 in][]{mt24}.

The four brightest datapoints inherent to Liu-Pang 1738 lacked viable or any Gaia DR3 XP results, and consequently they were manually dereddened using the cluster median (i.e., $E(B-R)=0.81\pm0.04$ and $E(B-R)=1.03\pm0.04$ for Alicante 13 and Liu-Pang 1738, accordingly). Uncertainties conveyed for those datapoints in Fig.~\ref{fig-ccd} are the median absolute deviation linked to cluster stars. 

The increased reddening for Liu-Pang 1738 can be rationalized by its sightline deviating deeper into the foreground Cygnus Rift \citep[see Fig.~3 cont.~in][]{dt85}. The enhanced obscuration associated with the foreground Cygnus Rift begins near $\approx700-800$ pc \citep{str19}. Indeed, the $4^{\circ}$ separation between the unbound clusters is $\approx 160$ pc at their present heliocentric distance, which implies the progenitor was a giant molecular cloud \citep[e.g.,][]{hd15,ch23}.  

\subsection{NIR color-magnitude diagram}
$JH$ color-magnitude diagrams (Fig.~\ref{fig-cmd}) were compiled for Alicante 13 and Liu-Pang 1738 using 2MASS and UKIDSS-DR6 photometry \citep{luc08}.  UKIDSS-DR6 saturates toward brighter magnitudes and 2MASS covers such stars. A $J=12^{m}.5$ boundary was imposed between the surveys. UKIDSS-DR6 was standardized to the 2MASS system using common stars in the field.  Mean photometry for the classical Cepheid SV Vul was evaluated using literature data \citep{mon11,nge12,bre21}. 

Distances for Alicante 13 and Liu-Pang 1738 were determined by matching their $JH$ observations to the nearby benchmark cluster IC 2602 (turnoff/main-sequence). IC 2602 was chosen owing to its negligible reddening, and similar age to Alicante 13 \citep[e.g.,][]{can20}. The mean cluster reddening is $E(B_J-V)=0.03\pm0.01$, as inferred from \citet{mer91} $UB_JV$ photometry\footnote{$B_J$ and $B$ are Johnson and Gaia passbands, accordingly.} for B-stars unaffected by the Balmer discontinuity. The intrinsic relation stemmed from unpublished work by D.~T.~(coauthor).  A reddening slope of $E(U-B_J)/E(B_J-V)\approx0.72$ was adopted to determine $E(B_J-V)$ in this low obscuration case \citep{hj56}, and slope uncertainties are likewise relatively inconsequential as a result. The optical color-excess transforms to $E(J-H)\simeq 0.01$ for IC 2602 \citep[i.e., $E(J-H)/E(B_J-V)=0.31\pm0.06$,][]{maj16}. 

NIR passbands were selected to establish the cluster distances since extinction law uncertainties are less onerous in comparison to the optical (e.g., Gaia $BGR$), and deviant trends are mitigated \citep{maj16}. IC 2602 photometry stemmed from 2MASS, and proper motion bounds were adopted from \citet[][Hipparcos]{vl09}. The L21-DR3 distance for IC 2602 is $\pi =6.646 \pm 0.012$ mas ($d=150$ pc), which was used to anchor IC 2602 (i.e., $M_J$). That agrees with a result of $148.6\pm2.0$ pc from \citet[][]{vl09}.

A visual fit using the calibrated IC 2602 yielded constraints of $E(J-H)=0.20, 0.26(\pm0.03)$ and $d=2.30, 2.40(\pm0.13)$ kpc for Alicante 13 and Liu-Pang 1738, respectively. The distance uncertainty was evaluated via analytical and Monte Carlo approaches, which relied on propagating uncertainties associated with $R$ \citep[$R_{J,JH}=A_J/E(J-H)=2.18\pm0.06$,][]{maj16}, and color-excess and turnoff/main-sequence fitting. The fit emerged by shifting intrinsic data along the ordinate and abscissa to match observations, and the results are supported by the DR3 parallaxes and isochrones (\S \ref{sec:redundancies}, Table~\ref{table:parameters}). The $JH$ color-magnitude analysis (Fig.~\ref{fig-cmd}) and DR3 parallaxes consistently indicate Alicante 13 is nearer than Liu-Pang 1738 (Table~\ref{table:parameters}).  The average offset implied by the two methods is $\simeq 0.15$ kpc.

A solar Padova\footnote{\url{http://pleiadi.pd.astro.it/}} PARSEC 1.2S-YBC $\log{\tau}=7.45\pm0.10$ isochrone was deduced from a preliminary visual\footnote{Separate semi-automated fits using numerous Padova isochrones are described in \S \ref{sec:redundancies}.} fit to Alicante 13 $JH$ data, and is well constrained by evolved stars \citep[e.g., HD 339064,][]{neg20}. That is comparable to a mean $\log{\tau}=7.35\pm0.17$ inferred from Cepheid period-age relations \citep[$P\simeq45^{d}$ for SV Vul,][]{bon05,tur12,and16}. The choice of a solar isochrone is linked to $\rm{[Fe/H]}=0.11\pm0.10$ for SV Vul \citep[][]{sil23}.  Age estimates in the literature for Alicante 13 include: $\log{\tau}=7.44$ and $7.48$ \citep{can20,neg20}. Similarly, for Liu-Pang 1738 estimates include $\log{\tau}=7.341\pm0.287$ and $7.87$ \citep{dia21,can20}.  The results are summarized in Table~\ref{table:parameters}.

\citet{tb04} contested arguments that SV Vul is traversing the instability strip for the first time since it exhibits a decreasing period ($-214.3\pm5.5$ s yr$^{-1}$), which instead indicates it is a blueward 2$^{nd}$ or 4$^{th}$ crosser (their Fig.~2, and in \S2 they favor the 2$^{nd}$ crossing).  A first-crosser would display an increasing period, one that is two orders of magnitude greater than determined, and the disputed hypothesis is likewise  contradicted by the isochrone (Fig.~\ref{fig-cmd}). Berdnikov et al.~(in prep.) extend that original $dP/dt$ determination by two decades. The analysis indicates a decreasing period at a rate of $dP/dt=-231.2\pm1.4$ s yr$^{-1}$.

 \subsection{Redundancies}
 \label{sec:redundancies}
 An effort was made to mitigate confirmation bias \citep[e.g.,][and discussion therein]{fre24}, and consequently coauthor C.~B.~carried out a semi-independent $JH$ color-magnitude analysis following precepts outlined by \citet{bo19}.  That coauthor established the ensuing parameters for Alicante 13: $2.18\pm0.02$ kpc, $\tau=35\pm5$ Myr ($\log{\tau}=7.54$), and $E(J-H)=0.19\pm0.02$.  Furthermore, author (D.~M.) assembled a median absolute deviation and sigma-clip algorithm to semi-automatedly fit isochrones (Majaess et al., in prep.), which provides yet another alternative to visual fitting using IC 2602. Means ($\pm stdev$) inferred for Alicante 13 from PARSEC 1.2S-YBC, YBC+new Vega, OBC, PARSEC 2.0 base and separately with $\omega_i=0.4$ are: $\log{\tau}=7.53\pm0.05$, $d=2.30\pm0.12$ kpc, $E(J-H)=0.22\pm0.01$, and $\rm{[Fe/H]}=0\pm0.08$. For Liu-Pang 1738 the parameters are $\log{\tau}=7.60\pm0.04$, $d=2.49\pm0.11$ kpc, $E(J-H)=0.26\pm0.01$, and $\rm{[Fe/H]}=0.06\pm0.09$. The lack of later-type evolved members in Liu-Pang 1738 complicates the isochrone selection, yet the ($u-g$)/($g-r$) and differentially dereddened diagrams convey the clusters share similar turnoffs (Fig.~\ref{fig-ccd}). 
 
 Irrespective of whether the distances are anchored to IC 2602 (preferred), or inferred from numerous isochrone models, or defined by Gaia L21-DR3 parallaxes: the findings are consistent. 

\subsection{Is SV Vul overluminous?}
\label{sec:svvul}
\citet{neg20} demonstrated through cluster $G-(B-R)$ and $K_s-(J-K_s)$ color-magnitude diagrams that SV Vul was too bright relative to isochrones ($\Delta G \simeq0^{m}.9$ and $\Delta K_s\simeq0^{m}.6$).  Matching canonical isochrones to Cepheids and their crossings remains challenging \citep[e.g.,][]{xu04,bon05,bon24,and16}. Yet \citet{rie21} and \citet{owe22} noted that SV Vul was an outlier in period-magnitude relations when utilizing EDR3 parallaxes (i.e., too bright).

A potential overluminosity is bolstered by several indicators concurrently: isochrone mismatches relative to a $J-(J-H)$ color-magnitude diagram ($\Delta J \simeq0^{m}.5$, Fig.~\ref{fig-cmd}), and via a brightness inferred from both Alicante 13 and L21-DR3 distances which are offset from certain Cepheid period-magnitude relations. The latter optical to MIR Cepheid functions can favor a nearer SV Vul distance of $\simeq 1.8-2$ kpc ($\mu_0\simeq11.5$) \citep{ber00,nge12,maj13e}, whereas a brighter Cepheid is implied by the cluster distance of $2.30\pm0.13$ kpc ($\mu_0\simeq11.81$) and the L21-DR3 parallax for SV Vul ($2.488^{+0.137}_{-0.124}$ kpc, $\mu_0\simeq11.98$).

A consensus solution could not be reached that explains the suite of trends, and which simultaneously considers factors such as: the absence of a B-type companion for SV Vul in data from the International Ultraviolet Explorer, the Cepheid's comparatively low $\gamma$-velocity shifts \citep[$\Delta\gamma\approx2$ km s$^{-1}$,][]{eva15}, marginal differential reddening, potential mass-loss, ostensibly a second rather than fourth crosser, chemical abundances and certain evidence disfavoring rapid rotation based on turbulence and other considerations \citep[e.g.,][]{tb04,luc18}. Consequently, a canonical massive companion or rapid rotation may not explain SV Vul's brightness. For example, perhaps a fine-tuned early merger scenario can be invoked, or alternatively certain Galactic Cepheid calibrations and canonical isochrones both require refinement, or even cluster membership could be reconsidered pending velocities for numerous cluster stars and validated DR4 parallaxes. 

\section{Conclusions}
This study builds upon efforts to characterize SV Vul and its encompassing region \citep[e.g.,][]{tur84}, namely by employing new or unused data, such as deeper UKIDSS-DR6 photometry, fainter ultraviolet UVEX observations, and Gaia DR3 data (XP spectra, astrometry, photometry). A consistent distance to SV Vul and Alicante 13 is established by multiple means, whether anchored via IC 2602 ($d=2.30\pm0.13$ kpc), Padova isochrones, or L21-DR3 parallaxes (Table~\ref{table:parameters}). An inspection of the broader field indicates SV Vul and Alicante 13 may potentially belong to a sizable complex which likewise includes Liu-Pang 1738 (Fig.~\ref{fig-pm}), and the astrometry is comparable (e.g., $\sigma_\pi\simeq0.02$ mas, $\sigma_{\mu_\delta}\simeq0.06$ mas yr$^{-1}$, Table~\ref{table:parameters}).  Importantly, in unison UVEX color-color diagrams, Gaia XP dereddened color-magnitude plots, and isochrones indicate the clusters are likewise relatively coeval (\S \ref{sec:redundancies}, Fig.~\ref{fig-ccd}).

Going forward, an open question remains whether SV Vul is overluminous (\S \ref{sec:svvul}).  Indeed, published concerns surrounding the critical long-period Cepheids SV Vul and S Vul provide ample impetus to continue research. Second, a dynamical analysis of both clusters via radial velocities is desirable (comprehensive data presently unavailable), however, the velocity trend along the sightline ($2.0-2.5$ kpc) is comparatively flat. Such velocities could (in)validate certain hypotheses.  Lastly, additional related overdensities may exist in the broader field \citep[e.g.,][]{pal25}. 

\begin{acknowledgments}
This research relied on initiatives such as CDS, NASA ADS, arXiv, Hipparcos, UKIDSS, 2MASS, Padova isochrones, Gaia, UVEX, and kind staff at the Gaia Help Desk. TapVizieR and Gaia DR3 access were facilitated by python and Gemini 3. D.~Minniti is supported by ANID Fondecyt Regular grant No.~1220724, and by the BASAL Center for Astrophysics and Associated Tecnologies (CATA) through ANID grants ACE210002 and FB210003. M.~G.~acknowledges support from ANID FONDECYT Regular 1240755.  R.~K.~S.~acknowledges support from CNPq/Brazil through projects 308298/2022-5 and 421034/2023-8. For L.~N.~B.~the study was conducted under the state assignment of Lomonosov Moscow State University.
\end{acknowledgments}

\bibliography{article}{}
\bibliographystyle{aasjournalv7}

\end{document}